\documentclass[a4]{aaprep}

\usepackage{url}
\usepackage{graphicx}
\usepackage{natbib}
\bibpunct{(}{)}{;}{a}{}{,} % to follow the A&A style

\usepackage{txfonts}

\def\OO#1{{\cal O}(c^{-#1})}
\newcommand{\vecg}[1]{\mbox{\boldmath$#1$}}
\newcommand{\ve}[1]{\vecg{#1}}

\newcommand{\numb}[2]{{\left\{#1\right\}}_{#2}}
\newcommand{\unit}[2]{{\left[#1\right]}_{#2}}

\def\TDB{$TDB$}
\def\TCB{$TCB$}
\def\TCG{$TCG$}
\def\TAI{$TAI$}
\def\TT{$TT$}
\def\UTC{$UTC$}

\begin{document}

\title{Relativistic scaling of astronomical quantities
and the system of astronomical units}

\author{Sergei A. Klioner}

\institute{
Lohrmann Observatory, Dresden Technical University,
Mommsenstr. 13, 01062 Dresden, Germany}

\date{Received   / Accepted }
%\date{25 May 2006}

\abstract
{
For relativistic modelling of high-accuracy astronomical data several
time scales are used: barycentric and geocentric coordinate times,
\TCB\ and \TCG, as well as two additional time scales, \TDB\ and \TT\,
that are defined as linear functions of \TCB\ and \TCG, respectively.
}
{
The paper is devoted to a concise but still detailed explanation of the
reasons and the implications of the relativistic scalings of
astronomical quantities induced by the time scales \TDB\
and \TT.
}
{
We consequently distinguish between quantities and their numerical values expressed in some
units.
}
{
It is argued that the scaled time scales, the scaled spatial
coordinates and the scaled masses should be considered as distinct
quantities which themselves can be expressed in any units, and not as
numerical values of the same quantities expressed in some different,
non-SI units (``\TDB\ units'' and ``\TT\ units'').
Along the same lines of argumentation the system of astronomical units
is discussed in the relativistic framework. The whole freedom in the
definitions of the systems of astronomical units for \TCB\ and \TDB\ is
demonstrated. A number of possible ways to freeze the freedom are shown
and discussed. It is argued that in the future one should
think about converting AU into a defined quantity by fixing its value
in SI meters.
}
{}

\keywords{Astrometry -- Reference systems -- Relativity}

\titlerunning{Relativistic astronomical time scales}
\authorrunning{S.A.~Klioner}

\maketitle

\section{Introduction}

It is well known that the accuracy of modern astronomical observations
has attained a level where numerous relativistic effects can no longer be
ignored. Moreover, the whole set of astronomical concepts used for
interpretation of observational data has to be formulated in the
framework of general relativity. In recent years significant
progress has been achieved in this direction. A rigorous post-Newtonian
framework for relativistic data modelling has been adopted by the
International Astronomical Union \citep[ and reference
therein]{Soffel:et:al:2003}. Nevertheless, the situation is not yet
fully satisfactory. One of the main factors retarding the adoption
of a fully self-consistent relativistic framework for fundamental
astronomy is the existence of ``inertia'' or ``traditions'' which are
quite difficult to overcome. Some of these traditions are heavily based
on special approximations in the framework of Newtonian physics, some
other are based on a Newtonian-like interpretation of the theory of
relativity.

One of the controversial questions of the latter kind is the situation with the linear
scaling of astronomical time scales and spatial coordinates related to
the theory of relativity. Although this question is clear and even
almost trivial from the theoretical point of view, practical
implications of the scaling are sometimes tricky and often understood in
a confusing way. The aim of this paper is to provide a concise,
self-consistent and rigorous description of the whole situation with
relativistic scalings. Interestingly, the same discussion can be used
to clarify the definition of the system of astronomical units in the relativistic
framework. This subject has been only marginally discussed in the
literature and not all what was published on this subject was correct.

In Section \ref{Section:units-vs-numbers} the relations between
quantities and their numerical values expressed in some units are
summarized. The justification for and implications of \TDB, being a
scaled version of the coordinate time \TCB\ of the Barycentric
Celestial Reference System (BCRS), are discussed in Section
\ref{Section-TDB-TCB-BCRS}. Section \ref{Section:TT-TCG-GCRS} is
devoted to the relativistic scaling in the Geocentric Celestial
Reference System (GCRS). The concept of coordinate time scales is
elucidated in Section \ref{Section-proper-vs-coordinate}. The
difficulties that appear when using several scaled reference systems
are sketched in Section \ref{Section-mixing}. The system of
astronomical units in the Newtonian and relativistic frameworks is
discussed in Sections \ref{Section:AU-Newton} and
\ref{Section:AU-Einstein}, respectively. The practical usage of the
various scaled quantities and also astronomical units in the
relativistic context is given in Section \ref{Section:GME} with the
example of extracting the masses of the Sun and the Earth from DE405 in
SI units. The question of whether the astronomical units of
measurements are still needed in modern astronomical practice in their
current form is discussed in Section \ref{Section:AU-needed}.

\section{Quantities, their numerical values and units of measurements}
\label{Section:units-vs-numbers}

In order to discuss the scaling issues let us first clearly distinguish
between quantities and their numerical values which appear when the
quantities are expressed as numbers using some units of measurements.
According to \citet[][ definition 1.1]{VocabularyMetrology1993}, {\it
quantity is an attribute of a phenomenon, body or substance that may be
distinguished qualitatively and determined quantitatively}. {\it A
value (of a quantity)} is defined as {\it the magnitude of a particular
quantity generally expressed as a unit of measurement multiplied by a
number} \citep[][ definition 1.18]{VocabularyMetrology1993}. Thus, for
any quantity $A$ one has
\begin{equation}
\label{A=A[]A}
A=\numb{A}{XX}\, \unit{A}{XX},
\end{equation}
\noindent
where $\numb{A}{XX}$ is the numerical value (a pure number) of quantity
$A$ and $\unit{A}{XX}$ is the corresponding unit. Notations
$\numb{A}{}$ and $\unit{A}{}$ for the numerical value and unit of a
quantity $A$, respectively, are taken from the international standard
ISO\,31-0 \citep{ISO31-0}. Since in this paper we use several systems
of units, the subscript gives the name of the system of units. Below we
use index ``SI'' for the SI units \citep{SI2006} and index ``A'' for
the system of astronomical units discussed in Section
\ref{Section:AU-Newton} and \ref{Section:AU-Einstein} below.
When a relation is valid
with any system of units, as in the case of Eq. (\ref{A=A[]A}),
``$XX$'' is used.

The official metrological definition of the concept of ``unit'' is
given by \citet[][ definition 1.7]{VocabularyMetrology1993}: {\it a unit
(of measurement) is a particular quantity, defined and adopted by
convention, with which other quantities of the same kind are compared
in order to express their magnitudes relative to that quantity.}
Loosely speaking, a unit is a recipe of how an observer can realize a
specific physical quantity called ``unit''. The observer can then express numerically
all other quantities of the same kind (those of the same physical
dimensionality) by comparing them to that specific quantity. The
complete set of definitions of the concepts of quantities, values,
units, systems of units, etc. can be found in
\citet{VocabularyMetrology1993}. A detailed discussion of these
concepts in the framework of general relativity is given by
\citet{Guinot:1997}.

Now let us consider two quantities $A$ and $B$ having the same physical
dimensionality and related by the following formula
derived in some theoretical way:
\begin{equation}
\label{B=KA}
B=F\,A,
\end{equation}
\noindent
$F$ being a pure number, that is a numerical, dimensionless
coefficient. This formula relates quantities $A$ and $B$ irrespective
of any considerations of units. To get a relation between numerical
values of $A$ and $B$ one has to use Eq. (\ref{A=A[]A}) on both sides
of (\ref{B=KA}). In particular, one has
\begin{equation}
\label{B[]=KA[]}
\numb{B}{XX}=F\,\numb{A}{XX}
\end{equation}
\noindent
if and only if one uses the {\it same} units for both $A$ and $B$:
$\unit{A}{XX}=\unit{B}{XX}$. Note that one could start this discussion
with any kind of formula relating $A$ and $B$ with the same conclusion:
such a formula is also valid for numerical values of the quantities if
and only if the same units are used for both of them.

%In metrological terms Eq. (\ref{B=KA}) belongs to the calculus of quantities.

%%%%%%%%%%%%%%%%%%%

\subsection{Units of measurements vs. units of graduation}
\label{Section-measurement-vs-graduation}

Strictly speaking, the concept of ``units of measurements'' can be only
applied to measurable (observable) quantities (e.g. proper time), but
not to non-measurable (i.e. coordinate-dependent) quantities in the
framework of general relativity \citep{Guinot:1997}. For the latter
kind of quantities one introduces the concept of ``units of
graduation'', which is an alias of ``units of measurement'' for
non-measurable quantities. The concept of ``units of graduation'' was
introduced to stress that the quantity under consideration is not
measurable so that its ``unit'' cannot be {\sl directly} realized by
some physical measurements. However, it seems appropriate to
ignore this subtle semantic difference in astronomical literature.

Indeed, let us consider the theoretical formula relating the proper time $\tau$
of an observer with the coordinate time $t$ of some relativistic
reference system
\begin{equation}
\label{tau-t}
\tau=f(t).
\end{equation}
\noindent
Proper time $\tau$ is a measurable quantity while coordinate time $t$
is not. This equation can be derived from the metric tensor of the
corresponding reference system and from the trajectory of the observer
in that reference system. Clearly, this is a relation between
quantities $t$ and $\tau$ and has nothing to do with units to be used
at the next step to express these quantities as numbers. The same units
should be again used for both $\tau$ and $t$ if relation (\ref{tau-t})
is to be valid also for the numerical values of these two quantities.
If one decides to use the SI second as unit of measurement for proper
time $\tau$ and if one assumes (\ref{tau-t}) to be valid for the
numerical values of $\tau$ and $t$, the corresponding unit of
graduation of $t$ is ``SI-second-compatible unit of graduation''. It is
safe, however, to simply call that latter unit of graduation ``SI
second''. It seems unnecessary to distinguish ``measurable'' and
``non-measurable'' quantities at the level of units (a discussion of
this point can be found in \citep{Guinot:1997}): it is sufficient to
distinguish these two kinds of quantities at the level of their
physical meaning and properties. In the following we will call both
units of measurements and units of graduation just ``units''.

%%%%%%%%%%%%%%%%%%%%

\subsection{On the possibility of scaled units}
\label{section-scaled-units}

Let us note that a linear relation like Eq. (\ref{B=KA}) could be in
principle interpreted as a relation between numerical values of one and
the same quantity expressed in different units (one quantity $C$, two
different units related as $\unit{C}{1}=F^{-1}\,\unit{C}{2}$, so that
the corresponding numerical values $\numb{C}{1}=F\,\numb{C}{2}$, and in
Eq. (\ref{B=KA}) $B\equiv\numb{C}{1}$ and $A\equiv\numb{C}{2}$).
However, it is dangerous and confusing to introduce several units for
the same physical dimensionality (especially, if these units are so
close to each other that there is a possibility of confusion). The way
to introduce two different units is against the usual metrological
rules (one unit for one dimension) and also against the IAU Resolutions
1991 (Recommendation II) that recommended the use of the SI units for
all quantities appearing in astronomical coordinate systems (in
particular, the use of the SI second for all time scales). Section
\ref{Section-proper-vs-coordinate} below contains a further discussion
of the topic in connection with the concept of coordinate time scales.

%\newpage

\section{Relativistic scaling in the BCRS}
\label{Section-TDB-TCB-BCRS}

\subsection{Dynamical equations in the BCRS}

Let $(t=TCB,x^i)$ be the coordinate time and spatial coordinates of the
Barycentric Celestial Reference System of the IAU
\citep{IAU:2001,Rickman:2001,Soffel:et:al:2003}. From the BCRS metric
tensor one can derive the so-called Einstein-Infeld-Hoffmann (EIH) equations
of motion of massive bodies considered as mass monopoles with masses ${\cal M}_A$
(capital latin indices $A$, $B$ and $C$ enumerate the bodies):
\begin{eqnarray}\label{mono-aEi}
&&
\ddot{\ve{x}}_A=-\sum_{B\ne A} \mu_B\,
{\ve{r}_{AB}\over |\ve{r}_{AB}|^3}
\nonumber\\
&&+{1\over c^2}\,\sum_{B\ne A} \mu_B\,
{\ve{r}_{AB}\over |\ve{r}_{AB}|^3}\,
\Biggl\{
%(2\beta-1)
\sum_{C\ne B} {\mu_C\over |\ve{r}_{BC}|}
%%\nonumber\\
%%&&\phantom{+{1\over c^2}\,\sum_{B} \mu_B\,
%%{\ve{r}_{AB}\over |\ve{r}_{AB}|^3}\,
%%\Biggl\{ }
%+2(\gamma+\beta)
+4\,
\sum_{C\ne A} {\mu_C\over |\ve{r}_{AC}|}
%%\nonumber \\
%%&&
%%\phantom{
%%+{1\over c^2}\,\sum_{B} \mu_B\,
%%{\ve{r}_{AB}\over |\ve{r}_{AB}|^3}\,
%%\Biggl\{ }
+{3\over 2} {{\left(\ve{r}_{AB}\cdot \dot{\ve{x}}_B\right)}^2\over |\ve{r}_{AB}|^2}
\nonumber\\
&&
\phantom{
+{1\over c^2}\,\sum_{B} \mu_B\,
{\ve{r}_{AB}\over |\ve{r}_{AB}|^3}\,
\Biggl\{ }
-{1\over 2}\sum_{C\ne A,B} \mu_C\,
{\ve{r}_{AB}\,\cdot\,\ve{r}_{BC}\over |\ve{r}_{BC}|^3}
\nonumber\\
&&
\phantom{
+{1\over c^2}\,\sum_{B} \mu_B\,
{\ve{r}_{AB}\over |\ve{r}_{AB}|^3}\,
\Biggl\{ }
%-(1+\gamma)
-2
\,\dot{\ve{x}}_B\,\cdot\,\dot{\ve{x}}_B
%-\gamma\,
-
\dot{\ve{x}}_A\,\cdot\,\dot{\ve{x}}_A
%%\nonumber \\
%%&&
%%\phantom{
%%+{1\over c^2}\,\sum_{B} \mu_B\,
%%{\ve{r}_{AB}\over |\ve{r}_{AB}|^3}\,
%%\Biggl\{ }
%+2(1+\gamma)
+4
\,\dot{\ve{x}}_A\,\cdot\,\dot{\ve{x}}_B
\Biggr\}
\nonumber\\
&&+{1\over c^2}\,\sum_{B\ne A} \mu_B\,
{\dot{\ve{x}}_A-\dot{\ve{x}}_B
\over |\ve{r}_{AB}|^3}\,
\biggl\{
% 2(1+\gamma)
 4
\,\dot{\ve{x}}_A\,\cdot\,\ve{r}_{AB}
%-(2\gamma+1)
-3
\,\dot{\ve{x}}_B\,\cdot\,\ve{r}_{AB}\biggr\}\,
\nonumber\\
&&-{1\over c^2}\,
%\left(2\gamma+{3\over 2}\right)
{7\over 2}
\,\sum_{B\ne A}{\mu_B\over |\ve{r}_{AB}|}\,
\sum_{C\ne A,B} \mu_C\,{\ve{r}_{BC} \over |\ve{r}_{BC}|^3}
+\OO4,
\end{eqnarray}
\noindent
and the following equation for the time of
light propagation between two points $\ve{x}_1$ and $\ve{x}_2$ (again
for the solar system considered as a system of mass monopoles)
\begin{eqnarray}\label{Shapiro}
c\,(t_2-t_1)&=&|\ve{x}_2-\ve{x}_1|
\nonumber\\
&&
+
\sum_A{2\,\mu_A\over c^2}\,\ln\,{|\ve{r}_{1A}|+|\ve{r}_{2A}|+|\ve{r}_{21}|\over
|\ve{r}_{2A}|+|\ve{r}_{1A}|-|\ve{r}_{21}|}
+\OO4,
\end{eqnarray}
\noindent
where $\mu_A=G\,{\cal M}_A$, $G$ is the Newtonian gravitational
constant, $c$ is the light velocity, $\ve{x}_A$ is the position of body
$A$, a dot denotes the time derivative with respect to \TCB, and for
any two indices $\ve{r}_{AB}=\ve{x}_A-\ve{x}_B$. Quantities
$\mu_A=G\,{\cal M}_A$ are called ``mass parameters'' below in order to
clearly distinguish them from masses ${\cal M}_A$ and the gravitational
constant $G$. Eq. (\ref{mono-aEi}) was derived as early as in 1917 by
Lorentz and Droste and then re-derived by Einstein, Infeld and Hoffmann
from a more general point of view (see Section VII.C of
\citet{DSX:1991} for a detailed history of these equations). The EIH
equations have been used to construct accurate solar system ephemerides
starting from the middle of the 1970s. Eq. (\ref{Shapiro}) describes
the relativistic Shapiro time delay which is also well known since the
1960s and is widely used for astronomical data modeling. Note, that up
to some theoretical improvements, the BCRS was known already in the
1930s and even earlier. The IAU has only officially fixed the status
quo in the Resolution B1.3 (2000). In both equations above the
coordinate time $t=TCB$ of the BCRS is used. It is \TCB\ (and not \TDB\
or any other time scale) that was used since 1917 in all theoretical
works involving the aforementioned equations and the underlying
relativistic reference system.

\subsection{\TDB\ as a linear function of \TCB}

For the reasons of practical convenience one often uses
the so-called $t^*=\hbox{\TDB}$ which is a linear function of $t=\hbox{\TCB}$:
\begin{equation}
\label{TDB-TCB}
t^*=F\,t+t^*_0,
\end{equation}
\noindent
with $F=1-L_B$ and $t^*_0$ are defining constants adopted by the
\citet{IAU:2006}. The constants are fixed here in such a way that \TDB,
evaluated at the geocenter, remains as close as possible to \TT\ (see
Section \ref{Section:TT-TCG-GCRS} below). In particular, the mean rate
of \TDB\ coincides with the mean rate of
\TT\ when the transformation between \TDB\ and \TT\ is evaluated at the geocenter.
The mean rate of \TT\ in its turn is equal, to a high level of
accuracy, to the mean rate of the proper time of an observer situated
on the rotating geoid. \TT\ is directly available to the Earth-bound
observers through \TAI, \UTC\ or any other realizations of \TT. The
difference between \TDB\ and \TT\ does not exceed 0.002~s and can be
neglected for many applications. These circumstances and lower risk of
a damage if \TDB\ is confused with \TT\ (compared to possible damages
of confusing \TCB\ and \TT\ with their linear drift of about 0.5~s per
year) are the arguments usually put forward in favor of \TDB. Here we
use the new definition of \TDB\ recently adopted by the
\citet{IAU:2006}. The original IAU wording given in 1976 defines \TDB\
so ``that there be only periodic variations between these time-scales''
(\TT\ and \TDB). This definition is known to be fundamentally flawed since the resulting
\TDB\ is not a linear function of \TCB\ and cannot be used with usual
dynamical equations (\ref{mono-aEi})--(\ref{Shapiro})
\citep{Standish:1998,Soffel:et:al:2003}.

Two realizations of the \TDB\ widely used in practice are given by the
analytical formulas relating \TDB\ and \TT\ given by \citet{Moyer1981}
and \citet{FairheadBretagnon1990}. The former has a lower accuracy of
about 20~$\mu$s and contains only periodic terms as an attempt to
adhere to the IAU (1976) description of \TDB. The more accurate
formulation of \citet{FairheadBretagnon1990} contains many non-periodic
(polynomial and mixed) terms. This demonstrates that retaining only
periodic terms in the transformation between \TDB\ and \TT\ is only
possible as a numerical approximation for lower accuracies and shorter time spans
for which the analytical formulas should be valid.

Another time scale, very similar to \TDB\ and also linearly related to
\TCB, was described by \citet{Standish:1998} and is called $T_{eph}$.
The subtle difference between $T_{eph}$ and \TDB\ lies in the way the
constants in (\ref{TDB-TCB}) are chosen. For \TDB\ the constant
$L_B=1.550519768\times10^{-8}$ in (\ref{TDB-TCB}) is a defining one
while for $T_{eph}$ the constant $F$ is different for different
ephemerides and implicitly defined by the transformation between \TT\
and $T_{eph}$ used during the construction of each particular
ephemeris. The adopted \TDB\ value of $L_B$ is based on the work of
\citet{IrwinFukushima:1999} and \citet{HaradaFukushima:2003}, and on
the IAU Resolution B1.9 (2000) defining \TT. The additive constant in
(\ref{TDB-TCB}) plays no role for the purposes of this paper and will
not be discussed here.

\subsection{Scaling of spatial coordinates and mass parameters}

If one uses $t^*=TDB$ instead of \TCB\ it is natural also to introduce
scaled spatial coordinates $\ve{x}^*$ and scaled mass parameters $\mu^*$ for each body
as
\begin{eqnarray}
\label{TDB-compatible-x}
\ve{x}^*&=&F\,\ve{x},
\\
\label{TDB-compatible-mu}
\mu^*&=&F\,\mu.
\end{eqnarray}
\noindent
These additional scalings allow one to retain exactly the same form of
the principal dynamical equations (\ref{mono-aEi}) and (\ref{Shapiro}). Quantities
$\ve{x}$ and $\mu$ are called \TCB-compatible quantities (or simply
\TCB\ quantities) representing spatial coordinates and mass parameters.
Quantities $\ve{x}^*$ and $\mu^*$ can be called \TDB-compatible
quantities (or simply \TDB\ quantities). Let us
make here several comments:
\begin{itemize}
\item[(1)] Physical mass of a body corresponds to $\mu$ ({\it not} to
$\mu^*$). Quantity $\mu$ does not depend on the kind of experiments
used to get it and on where the observer measuring it is situated.
Moreover, $\mu$ represents also the mass parameter of the corresponding body in
the Geocentric Celestial Reference System (GCRS) of the IAU
\citep{IAU:2001,Rickman:2001,Soffel:et:al:2003}. On the other hand, the
scaling factor between $\mu$ and $\mu^*$ is related to the fact that
most of accurate observations were until now performed from the surface
of [rotating] Earth. This made \TT\ (\TAI\ or \UTC) convenient or even
natural to parametrize observations. This will change as soon as
sufficiently large amount of high-accuracy observations is performed
from space vehicles. Therefore, $\mu^*$ can only be considered as some
ad hoc parameter convenient from some practical point of view.

\item[(2)] It is confusing to believe that $t$, $t^*$, $\ve{x}$,
$\ve{x}^*$, $\mu$, $\mu^*$ are not quantities, but just values in
different units: $t$, $\ve{x}$ and $\mu$
are numerical values expressed ``in SI units'', and $t^*$, $\ve{x}^*$ and
$\mu^*$ are ``in \TDB\ units''. As discussed in Section
\ref{section-scaled-units}, Eqs. (\ref{TDB-TCB}),
(\ref{TDB-compatible-x}) and (\ref{TDB-compatible-mu}) are relations
between six distinct quantities, and the question of units has not been
discussed at all here. One can, for example, consider (\ref{TDB-TCB})
and (\ref{TDB-compatible-x}) as relativistic coordinate transformations
from $(t,\ve{x})$ to $(t^*,\ve{x}^*)$ introducing another reference
system BCRS$^*$ distinct from BCRS. On the other hand, some ``\TDB\
units'', distinct to the SI units, imply, in particular, that the
``\TDB\ second'' is no longer SI second (and the number ``9192631770''
as appears in the definition of SI second \citep{SI2006} should be
explicitly changed to some other number in the definition of that ``\TDB\
second''). Those non-SI units have never been defined or discussed
seriously. Moreover, since there exists a consensus, also enforced by
the IAU Resolutions (1991), that \TT, \TCG, \TCB\ and related
quantities are all measured in the SI units (or in "SI-compatible units of
graduation"; see Section \ref{Section-measurement-vs-graduation})),
it is quite natural to extent this rule to \TDB\ and
\TDB-compatible quantities and to make the semantics symmetric and
clear. In Section \ref{Section-proper-vs-coordinate} below one can find
additional arguments against the concept of ``TDB units''.

\end{itemize}

\subsection{Further implications of the three scalings}

Eq. (\ref{TDB-compatible-x}) relating $\ve{x}$ and $\ve{x}^*$ is valid
for any distance used simultaneously with \TCB\ and \TDB. In
particular, the \TCB-compatible semi-major axis $a$ of a planet is related
to the corresponding \TDB-compatible semi-major axis $a^*$ as
\begin{equation}\label{a*-a}
a^*=F\,a.
\end{equation}
\noindent
In the same way Eq. (\ref{TDB-TCB}) also implies corresponding relations
between time intervals. In particular, the \TCB-compatible and \TDB-compatible
orbital periods of a planet are related as
\begin{equation}
\label{P*-P}
P^*=F\,P,
\end{equation}
\noindent
and the corresponding mean motions ($n=2\,\pi/P$, $n^*=2\,\pi/P^*$) as
\begin{equation}
\label{n*-n}
n^*=F^{-1}\,n.
\end{equation}
\noindent
Correspondingly, the third Keplerian law for a massless particle moving
in the field of a central body reads
\begin{equation}\label{3kep-TCB}
a^3\,n^2=\mu
\end{equation}
\noindent
for the \TCB-compatible $a$, $n$ and $\mu$, and
\begin{equation}\label{3kep-TDB}
a^{*3}\,n^{*2}=\mu^*
\end{equation}
\noindent
for the \TDB-compatible ones.

As discussed in Section \ref{Section:units-vs-numbers}, relations
(\ref{TDB-TCB})--(\ref{3kep-TDB}) are also valid for numerical values
of the corresponding quantities if the {\sl same} units are used for
the quantities appearing on both sides of these equations. If the units
used for \TCB- and \TDB-compatible quantities
are the same one has
\begin{eqnarray}
\label{TDB-TCB-[]}
&&\numb{t^*}{XX}=F\,\numb{t}{XX},
\\
\label{TDB-compatible-x-[]}
&&\numb{\ve{x}^*}{XX}=F\,\numb{\ve{x}}{XX},
\\
\label{TDB-compatible-mu-[]}
&&\numb{\mu^*}{XX}=F\,\numb{\mu}{XX},
\\
\label{a*-a-[]}
&&\numb{a^*}{XX}=F\,\numb{a}{XX},
\\
\label{P*-P-[]}
&&\numb{P^*}{XX}=F\,\numb{P}{XX},
\\
\label{n*-n-[]}
&&\numb{n^*}{XX}=F^{-1}\,\numb{n}{XX},
\\
\label{3kep-TCB-[]}
&&\numb{a}{XX}^3\,\numb{n}{XX}^2=\numb{\mu}{XX},
\\
\label{3kep-TDB-[]}
&&\numb{a^{*}}{XX}^3\,\numb{n^{*}}{XX}^2=\numb{\mu^*}{XX},
\end{eqnarray}
\noindent
where subscript ``$XX$'' denotes the name of any chosen system of
units. Those ``same'' units could be the SI units (SI seconds and SI
meters) as recommended by the IAU, but also any other system of units:
for example, astronomical units that are widely used in astronomy
during the last two centuries.

\section{Relativistic scaling in the GCRS}
\label{Section:TT-TCG-GCRS}

Let us consider the Geocentric Celestial Reference System (GCRS) of the
IAU with coordinates $(T=TCG,\ve{X})$. For the reasons discussed above
for \TDB, it is often convenient to introduce a scaled version of \TCG\
called $T^{**}=TT$. For current clock accuracies the mean rate of \TT\
is indistinguishable from to the mean rate of the proper time of an observer situated
at the rotating geoid. The difference comes from the tidal effects and
does not exceed $10^{-17}$ in the rate and 1 ps in the amplitude of periodic
terms. Again the scaling of time coordinate makes it convenient to
introduce the scaled versions of spatial coordinates and mass parameters:
\begin{eqnarray}
\label{TT-TCG}
T^{**}&=&L\ T,
\\
\label{X**-X}
\ve{X}^{**}&=&L\,\ve{X},
\\
\label{TT-masses-TCG}
\mu^{**}&=&L\,\mu,
\end{eqnarray}
\noindent
with $L=1-L_G$, $L_G\equiv6.969290134\times10^{-10}$ being a defining
constant \citep{IAU:2001}. As discussed above $\mu$ is the same in both
GCRS and BCRS. Here again we argue that one should speak about six
independent quantities (three \TCG-compatible quantities $T$, $\ve{X}$ and
$\mu$ and three \TT-compatible ones $T^{**}$, $\ve{X}^{**}$ and $\mu^{**}$)
without any reference to units. The expression ``TT units'' is confusing
for the same reasons as ``TDB units'' and should be avoided. Numerical
values of these six quantities expressed in the same units are scaled
in the same way as the quantities themselves. For example, in the
SI units one has
\begin{eqnarray}
\label{TT-TCG-unit}
\numb{T^{**}}{SI}&=&L\ \numb{T}{SI},
\\
\label{X*-X-unit}
\numb{\ve{X}^{**}}{SI}&=&L\,\numb{\ve{X}}{SI},
\\
\label{TT-masses-TCG-unit}
\numb{\mu^{**}}{SI}&=&L\,\numb{\mu}{SI}\ .
\end{eqnarray}

\section{On the concept of coordinate time scales}
\label{Section-proper-vs-coordinate}

The question if and in which sense the SI second can be used with the coordinate time
scales is closely related to the understanding of what coordinate time
scales are. Although the concept of a coordinate time is crystal clear
for people trained in relativity, coordinate time scales maybe
sometimes very confusing for people using ``Newtonian common
sense''. In the literature one can sometimes meet wrong statements
about relativistic coordinate time scales. Among these wrong statements
one can find: (1) \TCB\ is the time in the barycenter of the solar
system, (2) \TCG\ is the time at the geocenter, (3) \TT\ is the time on
the rotating geoid, (4) an ideal clock put in these three locations
would keep \TCB, \TCG\ and \TT, respectively, (5) for \TDB\ no location
could be found where an ideal clock would keep it and this implies
``TDB seconds''. All these statements originate in an incomplete or
inconsistent knowledge of relativity resulting in a latent yearning to
save Newtonian absolute time and at least some of its nice features.
Let us try to depict the role of coordinate time scales.

Let us consider first proper time of an observer.
Since the SI second plays an important role for this discussion let us
cite its definition \citep{SI2006}:

{\it
The second is the duration of 9\,192\,631\,770 periods of the radiation
corresponding to the transition between the two hyperfine levels of the
ground state of the caesium 133 atom.
}

It is important to understand that, in full agreement with general
relativity, this definition contains no indication of any specific
location, gravitational potential or state of motion of the observer
realizing the SI second. This means that this definition is a {\it
recipe} how any observer can realize the SI second. For all observers
this recipe is the same and in this sense the SI second is the same for
all observers. This means also that an observer has no chance to notice
his motion and position looking only at the readings of its clock.
Therefore, the SI second by itself is the unit of {\it proper time}.

The differences between proper times of two different observers can
only be noticed when a comparison procedure for two clocks having
different trajectories is established. If these observers are located
at the same place at the moments of comparison (as, e.g., in the twin
paradox), the comparison procedure is obvious. Otherwise (which is the
typical case) it involves some (arbitrary) relativistic reference
system which defines the coordinate simultaneity. The concept of
simultaneity is an indispensable part of any clock comparison
algorithm. The coordinate simultaneity is the only logical possibility
to save the concept of simultaneity in general relativity
\citep{Klioner1992,PetitWolf2005}: two events are called simultaneous
if and only if the chosen coordinate time has the same value for both
of them.

Let us turn to the coordinate time scales \TCB, \TDB, \TCG\ and \TT.
These time scales are part of the mathematical model of space-time used
in general relativity. The mathematical model of spaced-time is called
reference system and represent a 4-dimensional chart allowing one to
assign four numerical labels for each space-time event. Three of these
four labels are called spatial coordinates and the fourth label is
called coordinate time. Coordinate time scales \TCB, \TDB, \TCG\ and
\TT\ are defined for any space-time event within solar system and far
beyond. All these coordinate time scales are coordinates and,
therefore, cannot be directly measured. They can only be {\it computed}
from the readings of some real clock(s). For this computation one
should use a theoretical relation between the proper and coordinate
time scales that follows from the basic principles of general
relativity. That theoretical relation involves certain model of the
solar system: the trajectory of the observer, the trajectories of the
massive bodies, their mass parameters, etc. For \TCB\ and \TDB\ this
model is given in the BCRS, and for \TCG\ and \TT\ in the GCRS.

Sometimes, especially for didactic reasons, it is useful to consider
a special imaginary observer (that is, a special trajectory for an
imaginary observer), the proper time of which coincides with the
considered coordinate time along the observer's trajectory. For example,
for \TCB\ such an observer is situated infinitely far from the solar
system (so that the gravitational potential of the solar system
vanishes at his location) and is at rest relative to the solar system
barycenter. Certainly one can find also an analogous observer for \TDB:
take the same observer as for \TCB, but moving with a constant velocity
so that the Lorentz time dilation exactly compensates the rate
difference between \TCB\ and \TDB. Note however, that such
observers are only useful as an illustration and their existence should
be overestimated. First, these observer does not help to define \TCB\
or \TDB: the proper time of an observer is defined only on his
trajectory, while \TCB\ and \TDB\ are both defined everywhere in the
solar system. Second, these observers do not help to relate a real
clock moving within the solar system to \TCB\ and \TDB. Third, for
\TCG\ and \TT\ it is not possible to find such imaginary observers. One
often argues that \TT\ ``is defined on the rotating geoid''. This is
not true since \TT\ is a coordinate time scale and is defined for
any event in the solar system. However, the relation between the proper
time of an observer and \TT\ shows that the proper time of an observer
situated on the rotating geoid is close to \TT\ computed along his
trajectory, but only up to terms of order $10^{-17}$ and up to periodic
terms of an amplitude of about 1 ps. It is just a close agreement for a
particular trajectory and no more than that. \TT\ can be computed along
any other trajectory and can be related to the proper time of any other
observer in the solar system. Note also that the constant $L_G$ in the
definition of \TT\ is decoupled from the geoid and will not be changed
when the definition of the geoid is improved.

From the theoretical point of view the situation with the pair
\TCG-\TT\ is completely symmetric with the situation with the pair
\TCB-\TDB: \TT\ is a scaled version of \TCG, while \TDB\ is a scaled
version of \TCB. Both scalings have no physical meaning, but were
chosen from the considerations of convenience: to make the difference
between the proper time of an observer on the rotating geoid and these
two coordinate time scales evaluated along his trajectory as small as
possible.

From the metrological point of view, the theoretical relations between
the proper time and the coordinate times are relations between
quantities and are independent of the choice of units. The same is true
for the theoretical relations between the coordinate time scales
themselves. This means that any units can be used to convert these
quantities into numerical values. The SI second is usually used for the
proper time of any observer. If the theoretical formulas related the
proper time and coordinate times are used to compute the values of the
corresponding coordinate time starting from the values of proper time
in SI seconds, the values of the coordinate time are expressed also in
SI seconds (see, Section \ref{Section-measurement-vs-graduation}). The
same is true for the theoretical formulas relating coordinate time
scales with each other. In this situation some non-SI ``TDB units'' are
completely artificial and unnecessary.

\section{Mixing scaled BCRS and scaled GCRS}
\label{Section-mixing}

The scaling of BCRS and GCRS is obvious and simple to manage if only
one of these reference systems is used. In practice, however,
relativistic models often involve quantities defined in both reference
systems. Good examples here are models for VLBI and LLR, but it is also
the case for virtually all kinds of observations. For example, VLBI
model contains station coordinates and Earth orientation parameters
defined in the GCRS while the positions of sources and solar system
bodies (e.g. the Earth and the Sun) are defined in the BCRS. These
``mixed'' models are not invariant under the scalings
(\ref{TDB-TCB})--(\ref{TDB-compatible-mu}) and
(\ref{TT-TCG})--(\ref{TT-masses-TCG}). As a result the coefficients
$L_B$ and $L_G$ (and the constant $L_C\equiv 1-(1-L_B)/(1-L_G)\approx
L_B-L_G$) explicitly appear in the standard VLBI model \citep[][ Chapter
11]{IERS:2004}. This makes the models less transparent conceptually and
more difficult to understand and maintain. It should be stressed that
the scalings (and the corresponding coefficients) represent
non-physical, conventional changes of the BCRS and the GCRS and do not
appear in normal relativistic considerations.

Moreover, with the increasing importance of spacecraft observations,
the number of coordinate systems we have to deal with has proliferated.
For example, in order to study local physics (e.g. rotational motion)
of Mars, Moon or Mercury one could introduce GCRS-like planetocentric
reference systems in the vicinity of each planet. In particular, these
local planetocentric reference systems introduce their own coordinate
time scales. Consistent adherence to the idea of scaled coordinate
times having the same rate as \TT\ at the centers of mass of these
planets would require special scaling factors for each of these
reference systems. This would make the data reduction schemes
disastrously complicated and obscure.

In principle, it would be cleaner from the point of view of theoretical
purity and consistency not to introduce these scalings at all. It is
however clear that the considerations of convenience and a kind of
tradition weigh against the full use of the original non-scaled
versions of the BCRS and GCRS.

\section{The system of astronomical units in Newtonian framework}
\label{Section:AU-Newton}

The reason to introduce astronomical units of measurements in the 19th
century was the fact that the accuracy of positional (angular)
observations was much higher than the accuracy of determination of
distances (e.g. solar parallax). Before the invention of radar and laser
ranging and related techniques it was much easier to measure the period
of motion of a planet than to determine the distance to that planet
from the Sun or from the Earth (only a kind of geometrical
triangulation could be used: e.g. observations of Venus transits or of
Eros in its close approach to the Earth). For that reason, solar
system ephemerides have always been first constructed in the so-called
astronomical units to use the full precision of positional observations and
only later (and only if necessary) were they converted into other units
directly available in a laboratory (e.g., metric units). The precision
of that last conversion could be [much] lower than the precision of the
ephemeris in astronomical units. The ephemeris in astronomical units is
sufficient, however, to predict angular positions of the bodies on the
sky.

Let us first forget about relativity and consider the classical Newtonian
situation. The system of astronomical units consists of three units:
one for time $t$, one for mass ${\cal M}$ and one for length $x$. From
now on we designate these astronomical units as $\unit{t}{A}$,
$\unit{{\cal M}}{A}$ and $\unit{x}{A}$, while the corresponding SI
units are $\unit{t}{SI}=$~second, $\unit{{\cal M}}{SI}=$~kilogram and
$\unit{x}{SI}=$~meter. The corresponding numerical values in
astronomical units are denoted as $\numb{t}{A}$, $\numb{{\cal M}}{A}$
and $\numb{x}{A}$, and in SI units $\numb{t}{SI}$, $\numb{{\cal
M}}{SI}$ and $\numb{x}{SI}$.

The astronomical unit of time is the day. The day is directly related to the SI second:
\begin{equation}\label{day}
\unit{t}{A}= {\rm day}\ =d\ \unit{t}{SI},
\end{equation}
\noindent
where $d=86400$ is a pure number. The astronomical unit of mass
is fixed to coincide with the ``solar mass''  (SM)
\begin{equation}\label{solar-mass}
\unit{\cal M}{A}= {\rm SM}\ = \alpha\,\unit{{\cal M}}{SI},
\end{equation}
\noindent
where $\alpha$ is a pure number giving the solar mass in SI kilograms.
The value of $\alpha$ should be determined from observations (see
below). The astronomical unit of length $\unit{x}{A}$ is called
``Astronomical Unit'' (AU)
\begin{equation}\label{AU}
\unit{x}{A}= {\rm AU}\ =\chi\,\unit{x}{SI},
\end{equation}
\noindent
where $\chi$ is the number of SI meters in one AU. The AU is defined
in a tricky way with no relation to $\chi$. First, one fixes the value of the Newtonian
gravitational constant $G$ expressed in astronomical units to coincide
with the value determined by Gauss in 1809 from a series
of observations available to him. For historical reasons that value is
used up to now as a defining constant in the definition of the system
of astronomical units ($k$ is the well-known and widely-used standard
notation for $\sqrt{\numb{G}{A}}$):
\begin{eqnarray}\label{k^2}
\numb{G}{A}\equiv k^2&=&0.01720209895^2
\nonumber\\
&=&0.0002959122082855911025.
\end{eqnarray}
\noindent
Clearly, in any system of units the dimensionality of $G$ is
${\unit{x}{XX}}^3\,\unit{t}{XX}^{-2}\,{\unit{{\cal M}}{XX}}^{-1}$ (and
in particular, in astronomical units the dimensionality is ${\rm AU}^3
{\rm day}^{-2} {\rm SM}^{-1}$). The AU is then defined to be the unit
of length with which the gravitational constant $G$ takes the numerical
value (\ref{k^2}). This definition of the astronomical unit was adopted
by the IAU in 1938. One can also say that the AU is the semi-major axis of
the [hypothetical] orbit of a massless particle which has exactly a
period of ${2\pi\over k}\approx365.256898326328\dots$ days
(astronomical units of time) in the framework of unperturbed Keplerian
motion around the Sun having the mass of 1 {\rm SM}
\citep{BrouwerClemence:1961,Standish:2005a}. Kepler's third law gives
\begin{equation}\label{3kep-[]-astr}
\numb{a}{A}^3\,\numb{n}{A}^2=\numb{G}{A}\,\numb{{\cal M}}{A},
\end{equation}
\noindent
where $\numb{a}{A}$ and $\numb{n}{A}$ are the numerical values of the
semi-major axis and mean motion of a Keplerian orbit expressed in
astronomical units of length and time, respectively, and $\numb{{\cal
M}}{A}$ is the mass of the central body in the astronomical units of
mass SM. Normally, in the classical Newtonian case for the mass of the
Sun ${\cal M}_\odot$ one can just put $\numb{{\cal
M}_\odot}{A}\equiv1$. Note, however, that SM is a unit, while ${\cal
M}_\odot$ is a quantity. Moreover, SM must not coincide with the real
physical mass ${\cal M}_\odot$ of the Sun, especially since the latter
is time-dependent (see Section \ref{Section:AU-needed}).

The system of astronomical units is defined by four numbers $d$,
$\alpha$, $\chi$ and $k$. In modern astronomical practice
\citep{Standish:2005a} the value of $\chi$ is determined from the whole
set of available observations (various ranging observations that
measure distances directly in SI units play here a crucial role). Then,
comparing (\ref{3kep-[]-astr}) and (\ref{3kep-TCB-[]}), the numerical value of
$\mu=G\,{\cal M}_\odot$ in SI units can be computed as
\begin{equation}\label{mu-[]-from-astr}
\numb{\mu}{SI}=k^2\,\numb{{\cal M}_\odot}{A}\ \chi^3\,d^{-2}.
\end{equation}
\noindent
The mass of the Sun in kg can then be computed by using the SI value for
$G$ ($\numb{G}{SI}=6.674\dots\times10^{-11}\ {\rm m}^3\,{\rm s}^{-2}\,{\rm
kg}^{-1}$), but this last step is not important
for precise work.

Using the relations between the astronomical and SI units one can write the
following relations between numerical values of time $t$, distances
(positions) $\ve{x}$ and mass parameters $\mu$
\begin{eqnarray}
\label{time-[]}
&&\numb{t}{A}=d^{-1}\,\numb{t}{SI},
\\
\label{x-[]}
&&\numb{\ve{x}}{A}=\chi^{-1}\,\numb{\ve{x}}{SI},
\\
\label{mu-[]}
&&\numb{\mu}{A}=d^2\,\chi^{-3}\,\numb{\mu}{SI},
\end{eqnarray}
\noindent
and for the period $P$, mean motion $n$ and semi-major axis $a$
of an orbit
\begin{eqnarray}
\label{P-[]}
&&\numb{P}{A}=d^{-1}\,\numb{P}{SI},
\\
\label{n-[]}
&&\numb{n}{A}=d\ \numb{n}{SI},
\\
\label{a-[]}
&&\numb{a}{A}=\chi^{-1}\,\numb{a}{SI}.
\end{eqnarray}

\section{The system of astronomical units in the relativistic framework}
\label{Section:AU-Einstein}

Up to recently, only \TDB\ was used as independent time argument of
modern ephemeris. In connection with efforts to construct new
ephemerides with \TCB\ (or to re-parametrize old ones) the system of
astronomical units in the relativistic framework has been considered
recently by several authors
\citep{Standish:1995,BrumbergSimon:2004,Standish:2005b,Pitjeva:2005b}.
Let us interpret here all the formulas in the previous Section as
formulas relating \TCB-compatible quantities. As we discussed in
Section \ref{Section-TDB-TCB-BCRS} the \TDB-compatible quantities are
related to the corresponding \TCB-compatible ones by a relativistic
scaling. Then one can introduce another ``\TDB-compatible'' system of
astronomical units (designated as ``$A*$''):
\begin{eqnarray}
\label{day*}
&&
\unit{t}{A*}= {\rm day}^*\ =d^*\,\unit{t}{SI},
\\
\label{solar-mass-*}
&&
\unit{{\cal M}}{A*}= {\rm SM}^*\ = \alpha^*\,\unit{{\cal M}}{SI},
\\
\label{AU-*}
&&
\unit{x}{A*}= {\rm AU}^*\ =\chi^*\,\unit{x}{SI},
\\
\label{k^2-*}
&&
\numb{G^*}{A*}= {\left(k^*\right)}^2,
\end{eqnarray}
\noindent
Let us first consider the four numbers
$d^*$, $\alpha^*$, $\chi^*$ and $k^*$
as arbitrary (totally independent of the corresponding four numbers
$d$, $\alpha$, $\chi$ and $k$ defining the \TCB-compatible
system of astronomical units) and write down the relations between numerical
values of \TCB-compatible quantities expressed in \TCB-compatible astronomical units
(e.g. $\numb{\mu}{A}$) and \TDB-compatible quantities expressed in \TDB-compatible
astronomical units (e.g. $\numb{\mu^*}{A*}$):
\begin{eqnarray}
\label{time-[]-A*}
&&\numb{t^*}{A*}=F\,\left({d^*\over d}\right)^{-1}\,\numb{t}{A},
\\
\label{x-[]-A*}
&&\numb{\ve{x}^*}{A*}=F\,\left({\chi^*\over\chi}\right)^{-1}\,\numb{\ve{x}}{A},
\\
\label{mu-[]-A*}
&&\numb{\mu^*}{A*}=F\,
  \left({d^*\over d}\right)^2
\,\left({\chi^*\over\chi}\right)^{-3}\,\numb{\mu}{A}.
\end{eqnarray}
\noindent
These relations should be compared to the corresponding relations
(\ref{TDB-TCB-[]})--(\ref{TDB-compatible-mu-[]}) in SI units.
Considering that $\numb{\mu^*}{A*}={\left(k^*\right)}^2\,
\numb{{\cal M}_\odot^*}{A*}$ and $\numb{\mu}{A}=k^2\,
\numb{{\cal M}_\odot}{A}$ one can rewrite Eq. (\ref{mu-[]-A*}) as
\begin{equation}
{{\left(k^*\right)}^2\,\numb{{\cal M}_\odot^*}{A*}\over k^2\,\numb{{\cal M}_\odot}{A}}\,
\left({\chi^*\over \chi}\right)^3\,
\left({d^*\over d}\right)^{-2}=F.
\end{equation}
\noindent
This is the only constraint on the involved constants. Starting from this
relation one can suggest many different ways to define both \TCB- and
\TDB-compatible systems of astronomical constants. One reasonable
additional consideration is that a ``day'' is defined to be 86400 seconds in
any time scale (in \TCB, \TDB, \TCG, \TT\ or proper time of an observer)
as was recently agreed by the IAU Working Group on Nomenclature in Fundamental Astronomy
\citep{Capitaine:2005}. This means that the physical duration of a day
depends on the time scale used. Therefore, ``day'' is defined by the
conversion factor 86400 and it is natural to put $d^*=d=86400$.
Considering this one has at least two choices:
\begin{itemize}
\item[I.] One can require that the solar mass is equal to 1 in
corresponding astronomical units in both cases and that the constants
$k^*$ and $k$ are equal as well \citep{Standish:1995}. This gives
$\numb{\mu^*}{A*}=\numb{\mu}{A}=k^2$ and together with $d^*=d$ leads to
\begin{equation}
\chi^*=F^{1/3}\,\chi.
\end{equation}
\noindent
This, in turn, gives
\begin{eqnarray}
\label{time-[]-A*-1}
&&\numb{t^*}{A*}=F\,\numb{t}{A},
\\
\label{x-[]-A*-1}
&&\numb{\ve{x}^*}{A*}=F^{2/3}\,\numb{\ve{x}}{A},
\\
\label{mu-[]-A*-1}
&&\numb{\mu^*}{A*}=\numb{\mu}{A}.
\end{eqnarray}
\noindent
Note the unusual scaling laws of distances expressed in astronomical units
and the astronomical unit itself in this case. These scaling laws have produced already
a lot of confusion in the literature.

\item[II.] Another possibility
\citep{BrumbergSimon:2004,Standish:2005b,Pitjeva:2005b}
is to retain the scaling laws of time, distance and mass in astronomical units
(\ref{time-[]-A*})--(\ref{mu-[]-A*}) in exactly the same form as
in SI units (\ref{TDB-TCB-[]})--(\ref{TDB-compatible-mu-[]}) and put
\begin{equation}
\chi^*=\chi.
\end{equation}
which together with $d^*=d$ gives
\begin{eqnarray}
\label{time-[]-A*-2}
&&\numb{t^*}{A*}=F\,\numb{t}{A},
\\
\label{x-[]-A*-2}
&&\numb{\ve{x}^*}{A*}=F\,\numb{\ve{x}}{A},
\\
\label{mu-[]-A*-2}
&&\numb{\mu^*}{A*}=F\,\numb{\mu}{A}\ .
\end{eqnarray}
\noindent
The only ``unusual'' consequence of this choice is that
${\left(k^*\right)}^2\,\numb{{\cal M}_\odot^*}{A*} \neq
k^2\,\numb{{\cal M}_\odot}{A}$. Since modern ephemerides constructed with
\TDB-compatible quantities use ${\left(k^*\right)}^2\,
\numb{{\cal M}_\odot^*}{A*}\equiv0.01720209895^2$ \citep{Standish:2005b}
this means that for \TCB-compatible units one has
$k^2\,\numb{{\cal M}_\odot}{A}=F^{-1}\,0.01720209895^2
\approx 0.00029591221287376846\dots$.
\end{itemize}

The second choice seems to be more natural.
Let us note finally, that contrarily to what can be inferred from some
publications, the definitions of astronomical units by no means
influence the relations between numerical \TCB- and \TDB-compatible
quantities (e.g. $\mu$ and $\mu^*$) in SI units: they remain to be
defined as shown in Eqs. (\ref{TDB-TCB-[]})--(\ref{n*-n-[]}).

It is unclear what role these ``scaled'' relativistic astronomical
units could play for new solar system ephemerides: astronomical units
are just units and one can use any definitions of them as long as the
definitions are known. It makes no sense just to reformulate the {\it
same} process to produce ephemerides with \TCB\ (instead of \TDB) and
\TCB-compatible astronomical units (instead of \TDB-compatible
astronomical units): the results after corresponding re-scaling must be
identical to the \TDB-compatible ones, provided that all the scalings are
performed in a consistent way. If all scaling factors appearing in the
process of ephemeris development were applied correctly
one can claim that with the {\it same} level of confidence one can just
re-scale an existing ephemeris constructed in \TDB\ into \TCB\
according to equations given above. The question of consistent use of
relativistic time scales in the process of constructing new solar
system ephemerides will be considered in detail elsewhere
\citep{Klioner:2007}.

%\section{Astronomical units and solar system ephemerides???}
%
%The only possible advantage of using directly \TCB\ to construct new
%ephemerides is simultaneous iterative improvement of the \TCB-\TT\
%relation. The observations made from the Earth surface are usually
%available in terms of \TT\ (\TAI\ or \UTC) while the dynamical
%equations of the planetary motion is parametrized by \TDB. Therefore,
%the ephemeris development requires a time transformation from \TT\ to
%\TDB. Since \TT\ and \TDB\ are fixed linear functions of \TCG\ and
%\TCB, respectively, \TCG\ and \TCB\ can be always used instead of \TT\
%and \TDB. Let us note, however, that (1) the same iterative improvement
%scheme could be also implemented with \TDB\ (and actually {\it is}
%implemented with \TDB\ since the employed \TT-\TDB\ relations are being
%improved with time: the \TT-\TDB\ relations used for newer ephemerides
%were constructed by using older ephemeridies), and (2) considering that
%the currently used \TT-\TDB\ relations \citep[see,
%e.g.,][]{IrwinFukushima:1999,HaradaFukushima:2003} are quite accurate
%it cannot be expected that the iterative scheme would significantly
%improve the final accuracy of the ephemerides: tiny changes in the
%\TCB-\TT\ relation produce even smaller changes in the ephemeris. The
%situation could change if other kinds of observational data which are
%particularly sensitive to timing information (e.g. pulsar timing) are
%used to produce new ephemerides.

\section{Numerical example: mass parameters, coordinates and velocities from DE405}
\label{Section:GME}

Let us illustrate how to extract numerical values of planetary mass parameters
in SI units from the existing ephemerides constructed using \TDB\ and
the corresponding system of astronomical units,
using as a specific example
JPL's DE405. In the header of DE405 one finds the following
\TDB-compatible numerical values:
\begin{eqnarray}
\label{chi*-DE405}
\chi^*&=&1.49597870691\times10^{11},
\\
\numb{\mu_\odot^*}{A*}&=&2.959122082855911025\times 10^{-4}\ .
\end{eqnarray}
\noindent
The latter number is just the Gaussian value of $k^2$ quoted in
(\ref{k^2}). This allows us to find from (\ref{mu-[]-from-astr})
\begin{eqnarray}
\label{GMSun-TDB-DE405}
\numb{\mu_\odot^*}{SI}&=&\numb{\mu_\odot^*}{A*}\left(\chi^*\right)^3\,86400^{-2}
%\nonumber\\
=1.32712440018\times10^{20},
\end{eqnarray}
\noindent
and for the \TCB-compatible mass parameter
\begin{eqnarray}
\label{GMSun-TCB-DE405}
\numb{\mu_\odot}{SI}&=&F^{-1}\,\numb{\mu_\odot^*}{SI}=1.32712442076\times10^{20}.
\end{eqnarray}
\noindent
That latter value for the solar mass parameter can be found, e.g., in
\citet{IERS:2004}. The \TDB- and \TCB-compatible masses of
planets can be found from their \TDB-compatible masses in astronomical units
given in the header of DE405 in the same way. For example, for the Earth
one has
\begin{eqnarray}
\label{GMEarth-TDB-DE405}
\numb{\mu_\oplus^*}{SI}&=&3.98600432889\times10^{14},
\\
\label{GMEarth-TCB-DE405}
\numb{\mu_\oplus}{SI} &=&F^{-1}\,\numb{\mu_\oplus^*}{SI}=3.98600439069\times10^{14}.
\end{eqnarray}
The corresponding \TT-compatible value is then
\begin{eqnarray}
\label{GMEarth-TT-DE405}
\numb{\mu_\oplus^{**}}{SI}&=&L\,\numb{\mu_\oplus}{SI}=3.98600438792\times10^{14}.
\end{eqnarray}

\noindent
Let us also note that the mass parameters of the major planets, except for
Pluto, are all based currently on observations of spacecraft motions in
the vicinity of the corresponding planet. This means that an additional
re-scaling procedure should be performed between the mass parameters used for
different applications (e.g., for the Earth
$\numb{\mu_\oplus^{**}}{SI}$ is used for SLR, $\numb{\mu_\oplus^*}{SI}$
for the planetary ephemerides etc.).

Let us now turn to positions and velocities. The DE data in the
distribution gives the numerical values of the TDB-compatible spatial
coordinates $\ve{x}^*$ in SI units, that is $\numb{\ve{x}^*}{SI}$
parametrized by $t^*$ (precisely speaking the coordinates are given in kilometers, not in
meters, but it plays here no role).
If the TCB-compatible positions $\ve{x}$
are desired in SI units they can be computed as (cf. Eq. (\ref{TDB-compatible-x-[]}))
\begin{equation}
\label{TDB-x-TCB}
\numb{\ve{x}}{SI}=F^{-1}\,\numb{\ve{x}^*}{SI}.
\end{equation}
\noindent
The TCB-compatible velocity coincides with the TDB-compatible one
$\ve{v}^*=d\ve{x}^*/dt^*=d\ve{x}/dt=\ve{v}$. The standard JPL software
can be asked to output either directly $\numb{\ve{x}^*}{SI}$ or the
\TDB-compatible position $\ve{x}^*$ in \TDB-compatible astronomical
units, that is $\numb{\ve{x}^*}{A*}$. The latter value is computed by
dividing $\numb{\ve{x}^*}{SI}$ by $\chi^*$ from the header of the
ephemerides (see, Eq. (\ref{chi*-DE405})). A similar procedure applies to
the velocity $\ve{v}^*$. Note that it is only $\chi^*$, and therefore,
only TDB-compatible astronomical units that can be considered as a part
of the DE ephemerides. The choice of the TCB-compatible astronomical
units (and in particular the value of $\chi$) is by no means influenced
by the DE ephemerides or by the procedures used during their
development. Let us stress again that this choice has no influence on
the relation between the values in SI units.

\section{Do we need astronomical units in their current form?}
\label{Section:AU-needed}

It is not clear if astronomical units should be further used to
construct future ephemerides. The main reason for astronomical units --
much higher accuracy of angular (positional) observations compared to
distance measurements -- does not exist any longer. Considering the
subtleties with astronomical units in the relativistic framework one
can find it more advantageous either to avoid astronomical units at all
or for reasons of historical continuity to convert them into {\sl
defined} units by fixing $\chi$ as it was done with the day
($d\equiv86400$) and with the SI second.

One more argument against the system of astronomical units in its
current form is that the physical mass of the Sun is, in principle, not
constant, but decreasing at a rate of $\sim 10^{-11}$ solar masses per
century \citep{Noerdlinger:1997,KrasinskiBrumberg:2004} just because of
the Solar radiation. Up to now the dynamical consequences of this
change were below the accuracy of observations, but one can expect that
in the near future astronomical measurements in the inner solar system
will reach a level of accuracy where the effects of changing solar mass
(secular acceleration in the mean longitudes of the planets) will become observable.
For example, \citet{Pitjeva:2005a} gives the accuracy of the
determination of $\dot G/G$ as $5\times 10^{-12}$ per century. This is
the precision of the claim that no secular accelerations in the mean
longitudes of the inner planets are observable. On the other hand, a
linear change of the mass of the Sun has the same consequences for
astronomical observations as a linear change of $G$. Thus, in the near
future we will have to decide if we want to live with time-dependent
units of length, fix some epoch to define the Astronomical Unit, avoid
astronomical units in precise work, or, preferably, make the
Astronomical Unit to be a defined constant by fixing $\chi$ for
historical continuity.

If the AU is fixed in SI meters, the mass of the Sun or, more
precisely, $\mu_\odot=GM_\odot$ should be fitted from observations
together with masses of other planets, while the AU plays the same role
of a ``convenient'' unit as kilometer or mile. It seems to be even more
reasonable since in modern practice the masses of the planets are often
determined by other kinds of observations that deliver $\mu$ directly
in SI units. For example, the current best value for $\mu_\oplus$ is
delivered by SLR \citep{Groten:1999,Ries:2005} with no relation to
astronomical units.

\acknowledgement

The author is indebted to E. Myles Standish and Elena Pitjeva for their
patience in explaining his numerous questions. The author is also
thankful to Nicole Capitaine, Bernard Guinot, George Kaplan, Gerard
Petit, John Ries, Ken Seidelmann, Michael Soffel and Patrick Wallace
for insightful discussions.
This work was partially supported by the BMWi grant 50\,QG\,0601
awarded by the Deutsche Zentrum f\"ur Luft- und Raumfahrt e.V. (DLR).

\end{document}